\documentclass[pre,eqsecnum,preprint,showkeys,preprintnumbers,amssymb,amsfonts]{revtex4}

\input epsf

\usepackage{graphicx}
\usepackage{bm}

\newcommand{\beq}{\begin{equation}}
\newcommand{\eeq}{\end{equation}}
\newcommand{\beqs}{\begin{eqnarray}}
\newcommand{\eeqs}{\end{eqnarray}}

\newtheorem{lemma}{Lemma}[section]
\newtheorem{defi}{Definition}[section]

\newtheorem{propo}{Proposition}[section]

\begin{document}

\title{Number of connected spanning subgraphs on the Sierpinski gasket}

\author{Shu-Chiuan Chang$^a$} 
\email{scchang@mail.ncku.edu.tw} 

\author{Lung-Chi Chen$^b$} 
\email{lcchen@math.fju.edu.tw}

\affiliation{(a) \ Department of Physics \\
National Cheng Kung University \\
Tainan 70101, Taiwan} 

\affiliation{(b) \ Department of Mathematics \\
Fu Jen Catholic University \\
Taipei 24205, Taiwan}


\begin{abstract}

We study the number of connected spanning subgraphs $f_{d,b}(n)$ on the generalized Sierpinski gasket $SG_{d,b}(n)$ at stage $n$ with dimension $d$ equal to two, three and four for $b=2$, and layer $b$ equal to three and four for $d=2$. The upper and lower bounds for the asymptotic growth constant, defined as $z_{SG_{d,b}}=\lim_{v \to \infty} \ln f_{d,b}(n)/v$ where $v$ is the number of vertices, on $SG_{2,b}(n)$ with $b=2,3,4$ are derived in terms of the results at a certain stage. The numerical values of $z_{SG_{d,b}}$ are obtained.


\end{abstract}

\maketitle

\section{Introduction}
\label{sectionI}

The enumeration of the number of connected spanning subgraphs $N_{CSSG}(G)$ on a graph $G$ is a problem of interest in mathematics \cite{liu}. It is well known that the number of connected subgraphs is given by the Tutte polynomial $T(G,x,y)$ evaluated at $x=1$, $y=2$ \cite{welsh}. Alternatively, it corresponds to the partition function of the $q$-state Potts model in statistical mechanics with $q=0$ and the temperature variable $v=e^{\beta J}-1=1$, where $J$ is the spin-spin coupling and $\beta=(k_B T)^{-1}$. Some recent studies on the enumeration of connected spanning subgraphs and the calculation of their asymptotic growth constants on regular lattices were carried out in Refs. \cite{a,ta,hca,ka,s3a}. It is of interest to consider connected spanning subgraphs on self-similar fractal lattices which have scaling invariance rather than translational invariance. Fractals are geometric structures of (generally noninteger) Hausdorff dimension realized by repeated construction of an elementary shape on progressively smaller length scales \cite{mandelbrot,Falconer}. A well-known example of a fractal is the Sierpinski gasket. We shall derive the recursion relations for the numbers of connected spanning subgraphs on the Sierpinski gasket with dimension equal to two, three and four, and determine the asymptotic growth constants. We shall also consider the number of connected spanning subgraphs on the generalized Sierpinski gasket with dimension equal to two.

\section{Preliminaries}
\label{sectionII}

We first recall some relevant definitions for connected spanning subgraphs and the Sierpinski gasket in this section. A connected graph (without loops) $G=(V,E)$ is defined by its vertex (site) and edge (bond) sets $V$ and $E$ \cite{bbook,fh}.  Let $v(G)=|V|$ be the number of vertices and $e(G)=|E|$ the number of edges in $G$.  A spanning subgraph $G^\prime$ is a subgraph of $G$ with the same vertex set $V$ and an edge set $E^\prime \subseteq E$. A connected spanning subgraph on $G$ is a spanning subgraph of $G$ that remains connected. In general, there can be cycles in a connected spanning subgraph. It is called a spanning tree when there is no cycles. The degree or coordination number $k_i$ of a vertex $v_i \in V$ is the number of edges attached to it.  A $k$-regular graph is a graph with the property that each of its vertices has the same degree $k$. In general, one can associate an edge weight $x_{ij}$ to each edge connecting adjacent vertices $v_i$ and $v_j$. For simplicity, all edge weights are set to one throughout this paper. 

When the number of connected spanning subgraphs $N_{CSSG}(G)$ grows exponentially with $v(G)$ as $v(G) \to \infty$, there exists a constant $z_G$ describing this exponential growth:
\beq
z_G = \lim_{v(G) \to \infty} \frac{\ln N_{CSSG}(G)}{v(G)} \ ,
\label{zdef}
\eeq
where $G$, when used as a subscript in this manner, implicitly refers to
the thermodynamic limit.

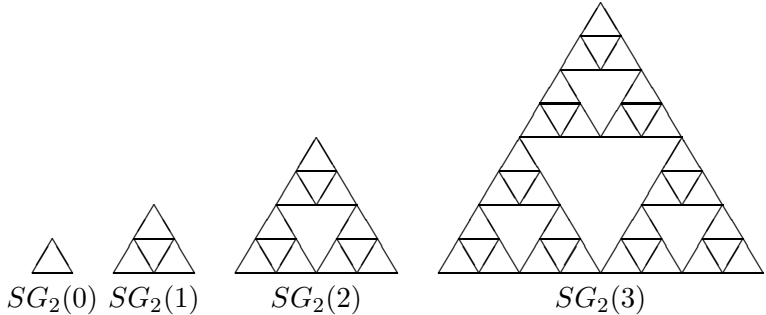
\begin{figure}[htbp]
\unitlength 0.9mm \hspace*{3mm}
\begin{picture}(108,40)
\put(0,0){\line(1,0){6}}
\put(0,0){\line(3,5){3}}
\put(6,0){\line(-3,5){3}}
\put(3,-4){\makebox(0,0){$SG_2(0)$}}
\put(12,0){\line(1,0){12}}
\put(12,0){\line(3,5){6}}
\put(24,0){\line(-3,5){6}}
\put(15,5){\line(1,0){6}}
\put(18,0){\line(3,5){3}}
\put(18,0){\line(-3,5){3}}
\put(18,-4){\makebox(0,0){$SG_2(1)$}}
\put(30,0){\line(1,0){24}}
\put(30,0){\line(3,5){12}}
\put(54,0){\line(-3,5){12}}
\put(36,10){\line(1,0){12}}
\put(42,0){\line(3,5){6}}
\put(42,0){\line(-3,5){6}}
\multiput(33,5)(12,0){2}{\line(1,0){6}}
\multiput(36,0)(12,0){2}{\line(3,5){3}}
\multiput(36,0)(12,0){2}{\line(-3,5){3}}
\put(39,15){\line(1,0){6}}
\put(42,10){\line(3,5){3}}
\put(42,10){\line(-3,5){3}}
\put(42,-4){\makebox(0,0){$SG_2(2)$}}
\put(60,0){\line(1,0){48}}
\put(72,20){\line(1,0){24}}
\put(60,0){\line(3,5){24}}
\put(84,0){\line(3,5){12}}
\put(84,0){\line(-3,5){12}}
\put(108,0){\line(-3,5){24}}
\put(66,10){\line(1,0){12}}
\put(90,10){\line(1,0){12}}
\put(78,30){\line(1,0){12}}
\put(72,0){\line(3,5){6}}
\put(96,0){\line(3,5){6}}
\put(84,20){\line(3,5){6}}
\put(72,0){\line(-3,5){6}}
\put(96,0){\line(-3,5){6}}
\put(84,20){\line(-3,5){6}}
\multiput(63,5)(12,0){4}{\line(1,0){6}}
\multiput(66,0)(12,0){4}{\line(3,5){3}}
\multiput(66,0)(12,0){4}{\line(-3,5){3}}
\multiput(69,15)(24,0){2}{\line(1,0){6}}
\multiput(72,10)(24,0){2}{\line(3,5){3}}
\multiput(72,10)(24,0){2}{\line(-3,5){3}}
\multiput(75,25)(12,0){2}{\line(1,0){6}}
\multiput(78,20)(12,0){2}{\line(3,5){3}}
\multiput(78,20)(12,0){2}{\line(-3,5){3}}
\put(81,35){\line(1,0){6}}
\put(84,30){\line(3,5){3}}
\put(84,30){\line(-3,5){3}}
\put(84,-4){\makebox(0,0){$SG_2(3)$}}
\end{picture}

\vspace*{5mm}
\caption{\footnotesize{The first four stages $n=0,1,2,3$ of the two-dimensional Sierpinski gasket $SG_2(n)$.}} 
\label{sgfig}
\end{figure}

The construction of the two-dimensional Sierpinski gasket $SG_2(n)$ at stage $n$ is shown in Fig. \ref{sgfig}. At stage $n=0$, it is an equilateral triangle; while stage $(n+1)$ is obtained by the juxtaposition of three $n$-stage structures. In general, the Sierpinski gaskets $SG_d$ can be built in any Euclidean dimension $d$ with fractal dimension $D=\ln(d+1)/\ln2$ \cite{Gefen81}. For the Sierpinski gasket $SG_d(n)$, the numbers of edges and vertices are given by 
\beq
e(SG_d(n)) = {d+1 \choose 2} (d+1)^n = \frac{d}{2} (d+1)^{n+1} \ ,
\label{e}
\eeq
\beq
v(SG_d(n)) = \frac{d+1}{2} [(d+1)^n+1] \ .
\label{v}
\eeq
Except the $(d+1)$ outmost vertices which have degree $d$, all other vertices of $SG_d(n)$ have degree $2d$. In the large $n$ limit, $SG_d$ is $2d$-regular. 

\bigskip

\bigskip

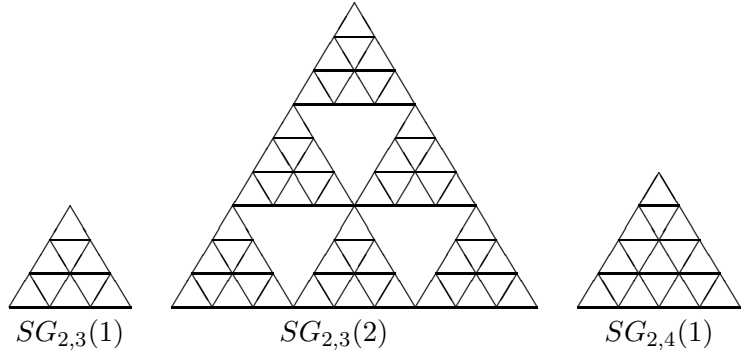
\begin{figure}[htbp]
\unitlength 0.9mm \hspace*{3mm}
\begin{picture}(108,45)
\put(0,0){\line(1,0){18}}
\put(3,5){\line(1,0){12}}
\put(6,10){\line(1,0){6}}
\put(0,0){\line(3,5){9}}
\put(6,0){\line(3,5){6}}
\put(12,0){\line(3,5){3}}
\put(18,0){\line(-3,5){9}}
\put(12,0){\line(-3,5){6}}
\put(6,0){\line(-3,5){3}}
\put(9,-4){\makebox(0,0){$SG_{2,3}(1)$}}
\put(24,0){\line(1,0){54}}
\put(33,15){\line(1,0){36}}
\put(42,30){\line(1,0){18}}
\put(24,0){\line(3,5){27}}
\put(42,0){\line(3,5){18}}
\put(60,0){\line(3,5){9}}
\put(78,0){\line(-3,5){27}}
\put(60,0){\line(-3,5){18}}
\put(42,0){\line(-3,5){9}}
\multiput(27,5)(18,0){3}{\line(1,0){12}}
\multiput(30,10)(18,0){3}{\line(1,0){6}}
\multiput(30,0)(18,0){3}{\line(3,5){6}}
\multiput(36,0)(18,0){3}{\line(3,5){3}}
\multiput(36,0)(18,0){3}{\line(-3,5){6}}
\multiput(30,0)(18,0){3}{\line(-3,5){3}}
\multiput(36,20)(18,0){2}{\line(1,0){12}}
\multiput(39,25)(18,0){2}{\line(1,0){6}}
\multiput(39,15)(18,0){2}{\line(3,5){6}}
\multiput(45,15)(18,0){2}{\line(3,5){3}}
\multiput(45,15)(18,0){2}{\line(-3,5){6}}
\multiput(39,15)(18,0){2}{\line(-3,5){3}}
\put(45,35){\line(1,0){12}}
\put(48,40){\line(1,0){6}}
\put(48,30){\line(3,5){6}}
\put(54,30){\line(3,5){3}}
\put(54,30){\line(-3,5){6}}
\put(48,30){\line(-3,5){3}}
\put(48,-4){\makebox(0,0){$SG_{2,3}(2)$}}
\put(84,0){\line(1,0){24}}
\put(87,5){\line(1,0){18}}
\put(90,10){\line(1,0){12}}
\put(93,15){\line(1,0){6}}
\put(84,0){\line(3,5){12}}
\put(90,0){\line(3,5){9}}
\put(96,0){\line(3,5){6}}
\put(102,0){\line(3,5){3}}
\put(108,0){\line(-3,5){12}}
\put(102,0){\line(-3,5){9}}
\put(96,0){\line(-3,5){6}}
\put(90,0){\line(-3,5){3}}
\put(96,-4){\makebox(0,0){$SG_{2,4}(1)$}}
\end{picture}

\vspace*{5mm}
\caption{\footnotesize{The generalized two-dimensional Sierpinski gasket $SG_{2,b}(n)$ with $b=3$ at stage $n=1, 2$ and $b=4$ at stage $n=1$.}} 
\label{sgbfig}
\end{figure}

The Sierpinski gasket can be generalized, denoted by $SG_{d,b}(n)$, by introducing the side length $b$ which is an integer larger or equal to two \cite{Hilfer}. The generalized Sierpinski gasket at stage $(n+1)$ is constructed from $b$ layers of stage $n$ hypertetrahedrons. The two-dimensional $SG_{2,b}(n)$ with $b=3$ at stage $n=1, 2$ and $b=4$ at stage $n=1$ are illustrated in Fig. \ref{sgbfig}. The ordinary Sierpinski gasket $SG_d(n)$ corresponds to the $b=2$ case, where the index $b$ is neglected for simplicity. The Hausdorff dimension for $SG_{d,b}$ is given by $D=\ln {b+d-1 \choose d} / \ln b$ \cite{Hilfer}. Notice that $SG_{d,b}$ is not $k$-regular even in the thermodynamic limit.

\bigskip

\bigskip

\section{The number of connected spanning subgraphs on $SG_2(n)$}
\label{sectionIII}

In this section we derive the asymptotic growth constant for the number of connected spanning subgraphs on the two-dimensional Sierpinski gasket $SG_2(n)$ in detail. Let us start with the definitions of the quantities to be used.

\bigskip

\begin{figure}[htbp]
\unitlength 1.8mm 
\begin{picture}(66,5)
\put(0,0){\line(1,0){6}}
\put(0,0){\line(3,5){3}}
\put(6,0){\line(-3,5){3}}
\put(3,-2){\makebox(0,0){$f_2(n)$}}
\put(12,0){\line(1,0){6}}
\multiput(12,0)(0.3,0.5){11}{\circle*{0.2}}
\multiput(18,0)(-0.3,0.5){11}{\circle*{0.2}}
\put(15,-2){\makebox(0,0){$g_2(n)$}}
\multiput(24,0)(0.5,0){13}{\circle*{0.2}}
\multiput(24,0)(0.3,0.5){11}{\circle*{0.2}}
\multiput(30,0)(-0.3,0.5){11}{\circle*{0.2}}
\put(27,-2){\makebox(0,0){$h_2(n)$}}
\end{picture}

\vspace*{5mm}
\caption{\footnotesize{Illustration for the connected spanning subgraphs $f_2(n)$, $g_2(n)$ and $h_2(n)$. The two outmost vertices at the ends of a solid line belong to one component, while the two outmost vertices at the ends of a dot line belong to separated components.}} 
\label{fghfig}
\end{figure}
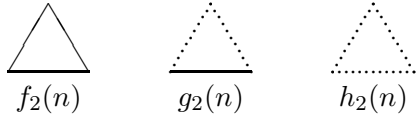

\begin{defi} \label{defisg2} Consider the generalized two-dimensional Sierpinski gasket $SG_{2,b}(n)$ at stage $n$. (i) Define $f_{2,b}(n) \equiv N_{CSSG}(SG_{2,b}(n))$ as the number of connected spanning subgraphs. (ii) Define $g_{2,b}(n)$ as the number of spanning subgraphs with two connected components such that one certain outmost vertex, say the topmost vertex as illustrated in Fig. \ref{fghfig} for ordinary Sierpinski gasket, belongs to one component and the other two outmost vertices belong to another component. (iii) Define $h_{2,b}(n)$ as the number of spanning subgraphs with three connected components such that each of the outmost vertices belongs to a different component.
\end{defi}

\bigskip

Since we only consider the ordinary Sierpinski gasket in this section, we use the notations $f_2(n)$, $g_2(n)$ and $h_2(n)$ for simplicity. They are illustrated in Fig. \ref{fghfig}, where only the outmost vertices are shown. Because of rotational symmetry, there are three possible $g_2(n)$. The initial values at stage zero are $f_2(0)=4$, $g_2(0)=1$ and $h_2(0)=1$. The purpose of this section is to obtain the asymptotic behavior of $f_2(n)$ as follows.
The three quantities $f_2(n)$, $g_2(n)$ and $h_2(n)$ satisfy recursion relations. 

\bigskip

\bigskip

\begin{lemma} \label{lemmasg2r} For any non-negative integer $n$,
\beq
f_2(n+1) = f_2^3(n) + 6f_2^2(n) g_2(n) \ , 
\label{feq}
\eeq
\beqs
g_2(n+1) & = & f_2^2(n)g_2(n) + f_2^2(n)h_2(n) + 7f_2(n)g_2^2(n) \ , 
\label{geq}
\eeqs
\beqs
h_2(n+1) & = & 3f_2(n)g_2^2(n) + 12f_2(n)g_2(n)h_2(n) + 14g_2^3(n) \ .
\label{heq}
\eeqs
\end{lemma}

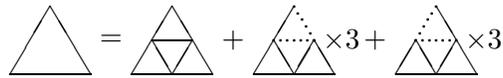
\begin{figure}[htbp]
\unitlength 0.9mm 
\begin{picture}(70,12)
\put(0,0){\line(1,0){12}}
\put(0,0){\line(3,5){6}}
\put(12,0){\line(-3,5){6}}
\put(15,5){\makebox(0,0){$=$}}
\put(18,0){\line(1,0){12}}
\put(21,5){\line(1,0){6}}
\put(18,0){\line(3,5){6}}
\put(30,0){\line(-3,5){6}}
\put(24,0){\line(3,5){3}}
\put(24,0){\line(-3,5){3}}
\put(33,5){\makebox(0,0){$+$}}
\put(36,0){\line(1,0){12}}
\multiput(39,5)(1,0){7}{\circle*{0.2}}
\put(36,0){\line(3,5){6}}
\put(42,0){\line(3,5){3}}
\put(42,0){\line(-3,5){3}}
\put(48,0){\line(-3,5){3}}
\multiput(45,5)(-0.6,1){6}{\circle*{0.2}}
\put(49,5){\makebox(0,0){$\times 3$}}
\put(54,5){\makebox(0,0){$+$}}
\put(57,0){\line(1,0){12}}
\multiput(60,5)(1,0){7}{\circle*{0.2}}
\put(57,0){\line(3,5){3}}
\put(69,0){\line(-3,5){6}}
\put(63,0){\line(3,5){3}}
\put(63,0){\line(-3,5){3}}
\multiput(60,5)(0.6,1){6}{\circle*{0.2}}
\put(70,5){\makebox(0,0){$\times 3$}}
\end{picture}

\caption{\footnotesize{Illustration for the expression of $f_2(n+1)$. The multiplication of three on the right-hand-side corresponds to the three possible orientations of $SG_2(n+1)$.}} 
\label{ffig}
\end{figure}

{\sl Proof} \quad 
The Sierpinski gasket $SG_2(n+1)$ is composed of three $SG_2(n)$ with three pairs of vertices identified. The number $f_2(n+1)$ consists of one configuration where all three $SG_2(n)$ belong to the class that is enumerated by $f_2(n)$, and six configurations where one of the $SG_2(n)$ belongs to the class enumerated by $g_2(n)$ and the other two belong to the class enumerated by $f_2(n)$ as illustrated in Fig. \ref{ffig}. Eq. (\ref{feq}) is verified by adding these configurations.

\bigskip

\bigskip

Similarly, $g_2(n+1)$ and $h_2(n+1)$ for $SG_2(n+1)$ can be obtained with appropriate configurations of its three constituting $SG_2(n)$ as illustrated in Figs. \ref{gfig} and \ref{hfig} to verify Eqs. (\ref{geq}) and (\ref{heq}), respectively. 

\bigskip

\begin{figure}[htbp]
\unitlength 0.9mm 
\begin{picture}(127,12)
\put(0,0){\line(1,0){12}}
\multiput(0,0)(0.6,1){11}{\circle*{0.2}}
\multiput(12,0)(-0.6,1){11}{\circle*{0.2}}
\put(15,5){\makebox(0,0){$=$}}
\put(18,0){\line(1,0){12}}
\put(21,5){\line(1,0){6}}
\multiput(18,0)(6,0){2}{\line(3,5){3}}
\multiput(24,0)(6,0){2}{\line(-3,5){3}}
\multiput(21,5)(0.6,1){6}{\circle*{0.2}}
\multiput(27,5)(-0.6,1){6}{\circle*{0.2}}
\put(33,5){\makebox(0,0){$+$}}
\put(36,0){\line(1,0){12}}
\multiput(39,5)(1,0){7}{\circle*{0.2}}
\multiput(36,0)(6,0){2}{\line(3,5){3}}
\multiput(42,0)(6,0){2}{\line(-3,5){3}}
\multiput(39,5)(0.6,1){6}{\circle*{0.2}}
\multiput(45,5)(-0.6,1){6}{\circle*{0.2}}
\put(51,5){\makebox(0,0){$+$}}
\put(54,0){\line(1,0){12}}
\put(57,5){\line(1,0){6}}
\put(57,5){\line(3,5){3}}
\put(63,5){\line(-3,5){3}}
\multiput(54,0)(0.6,1){6}{\circle*{0.2}}
\multiput(60,0)(0.6,1){6}{\circle*{0.2}}
\multiput(60,0)(-0.6,1){6}{\circle*{0.2}}
\multiput(66,0)(-0.6,1){6}{\circle*{0.2}}
\put(69,5){\makebox(0,0){$+$}}
\put(72,0){\line(1,0){12}}
\put(75,5){\line(1,0){6}}
\put(72,0){\line(3,5){3}}
\put(78,0){\line(-3,5){3}}
\multiput(78,0)(0.6,1){6}{\circle*{0.2}}
\multiput(75,5)(0.6,1){6}{\circle*{0.2}}
\multiput(84,0)(-0.6,1){11}{\circle*{0.2}}
\put(85,5){\makebox(0,0){$\times 2$}}
\put(90,5){\makebox(0,0){$+$}}
\put(93,0){\line(3,5){3}}
\multiput(93,0)(3,5){2}{\line(1,0){6}}
\multiput(99,0)(6,0){2}{\line(-3,5){3}}
\multiput(99,0)(1,0){7}{\circle*{0.2}}
\multiput(99,0)(0.6,1){6}{\circle*{0.2}}
\multiput(96,5)(0.6,1){6}{\circle*{0.2}}
\multiput(102,5)(-0.6,1){6}{\circle*{0.2}}
\put(106,5){\makebox(0,0){$\times 2$}}
\put(111,5){\makebox(0,0){$+$}}
\put(114,0){\line(1,0){12}}
\put(114,0){\line(3,5){3}}
\multiput(120,0)(3,5){2}{\line(-3,5){3}}
\multiput(120,0)(0.6,1){6}{\circle*{0.2}}
\multiput(117,5)(0.6,1){6}{\circle*{0.2}}
\multiput(117,5)(1,0){7}{\circle*{0.2}}
\multiput(126,0)(-0.6,1){6}{\circle*{0.2}}
\put(127,5){\makebox(0,0){$\times 2$}}
\end{picture}

\caption{\footnotesize{Illustration for the expression of  $g_2(n+1)$. The multiplication of two on the right-hand-side corresponds to the reflection symmetry with respect to the central vertical axis.}} 
\label{gfig}
\end{figure}

\bigskip

\begin{figure}[htbp]
\unitlength 0.9mm 
\begin{picture}(136,12)
\multiput(0,0)(1,0){13}{\circle*{0.2}}
\multiput(0,0)(0.6,1){11}{\circle*{0.2}}
\multiput(12,0)(-0.6,1){11}{\circle*{0.2}}
\put(15,5){\makebox(0,0){$=$}}
\multiput(18,0)(1,0){13}{\circle*{0.2}}
\multiput(18,0)(0.6,1){6}{\circle*{0.2}}
\multiput(30,0)(-0.6,1){6}{\circle*{0.2}}
\put(21,5){\line(1,0){6}}
\multiput(24,0)(3,5){2}{\line(-3,5){3}}
\multiput(24,0)(-3,5){2}{\line(3,5){3}}
\put(31,5){\makebox(0,0){$\times 3$}}
\put(36,5){\makebox(0,0){$+$}}
\multiput(39,0)(1,0){13}{\circle*{0.2}}
\multiput(39,0)(0.6,1){6}{\circle*{0.2}}
\multiput(45,0)(-0.6,1){6}{\circle*{0.2}}
\multiput(51,0)(-0.6,1){6}{\circle*{0.2}}
\put(42,5){\line(1,0){6}}
\put(48,5){\line(-3,5){3}}
\multiput(45,0)(-3,5){2}{\line(3,5){3}}
\put(52,5){\makebox(0,0){$\times 3$}}
\put(57,5){\makebox(0,0){$+$}}
\multiput(60,0)(1,0){13}{\circle*{0.2}}
\multiput(60,0)(0.6,1){6}{\circle*{0.2}}
\multiput(66,0)(0.6,1){6}{\circle*{0.2}}
\multiput(72,0)(-0.6,1){6}{\circle*{0.2}}
\put(63,5){\line(1,0){6}}
\put(63,5){\line(3,5){3}}
\multiput(66,0)(3,5){2}{\line(-3,5){3}}
\put(73,5){\makebox(0,0){$\times 3$}}
\put(78,5){\makebox(0,0){$+$}}
\multiput(81,0)(1,0){7}{\circle*{0.2}}
\multiput(81,0)(0.6,1){6}{\circle*{0.2}}
\multiput(87,0)(0.6,1){6}{\circle*{0.2}}
\multiput(87,0)(-0.6,1){6}{\circle*{0.2}}
\multiput(93,0)(-0.6,1){6}{\circle*{0.2}}
\put(84,5){\line(3,5){3}}
\put(90,5){\line(-3,5){3}}
\multiput(84,5)(3,-5){2}{\line(1,0){6}}
\put(94,5){\makebox(0,0){$\times 3$}}
\put(99,5){\makebox(0,0){$+$}}
\multiput(102,0)(3,5){2}{\line(1,0){6}}
\put(105,5){\line(3,5){3}}
\put(111,5){\line(-3,5){3}}
\multiput(108,0)(1,0){7}{\circle*{0.2}}
\multiput(102,0)(0.6,1){6}{\circle*{0.2}}
\multiput(108,0)(0.6,1){6}{\circle*{0.2}}
\multiput(108,0)(-0.6,1){6}{\circle*{0.2}}
\multiput(114,0)(-0.6,1){6}{\circle*{0.2}}
\put(115,5){\makebox(0,0){$\times 3$}}
\put(120,5){\makebox(0,0){$+$}}
\multiput(123,0)(6,0){2}{\line(3,5){3}}
\put(126,5){\line(1,0){6}}
\multiput(123,0)(1,0){13}{\circle*{0.2}}
\multiput(135,0)(-0.6,1){11}{\circle*{0.2}}
\multiput(126,5)(0.6,1){6}{\circle*{0.2}}
\multiput(129,0)(-0.6,1){6}{\circle*{0.2}}
\put(136,5){\makebox(0,0){$\times 3$}}
\end{picture}

\unitlength 0.9mm 
\begin{picture}(135,12)
\put(15,5){\makebox(0,0){$+$}}
\multiput(18,0)(1,0){13}{\circle*{0.2}}
\multiput(18,0)(0.6,1){11}{\circle*{0.2}}
\multiput(24,0)(0.6,1){6}{\circle*{0.2}}
\multiput(27,5)(-0.6,1){6}{\circle*{0.2}}
\put(21,5){\line(1,0){6}}
\multiput(24,0)(6,0){2}{\line(-3,5){3}}
\put(31,5){\makebox(0,0){$\times 3$}}
\put(36,5){\makebox(0,0){$+$}}
\put(39,0){\line(3,5){3}}
\multiput(45,0)(-3,5){2}{\line(1,0){6}}
\multiput(39,0)(1,0){7}{\circle*{0.2}}
\multiput(42,5)(0.6,1){6}{\circle*{0.2}}
\multiput(45,0)(-0.6,1){6}{\circle*{0.2}}
\multiput(45,0)(0.6,1){6}{\circle*{0.2}}
\multiput(51,0)(-0.6,1){11}{\circle*{0.2}}
\put(52,5){\makebox(0,0){$\times 3$}}
\put(57,5){\makebox(0,0){$+$}}
\multiput(60,0)(3,5){2}{\line(1,0){6}}
\put(72,0){\line(-3,5){3}}
\multiput(60,0)(0.6,1){11}{\circle*{0.2}}
\multiput(66,0)(0.6,1){6}{\circle*{0.2}}
\multiput(66,0)(-0.6,1){6}{\circle*{0.2}}
\multiput(66,0)(1,0){7}{\circle*{0.2}}
\multiput(69,5)(-0.6,1){6}{\circle*{0.2}}
\put(73,5){\makebox(0,0){$\times 3$}}
\put(78,5){\makebox(0,0){$+$}}
\put(81,0){\line(3,5){3}}
\put(87,0){\line(1,0){6}}
\put(90,5){\line(-3,5){3}}
\multiput(81,0)(1,0){7}{\circle*{0.2}}
\multiput(84,5)(1,0){7}{\circle*{0.2}}
\multiput(84,5)(0.6,1){6}{\circle*{0.2}}
\multiput(87,0)(0.6,1){6}{\circle*{0.2}}
\multiput(87,0)(-0.6,1){6}{\circle*{0.2}}
\multiput(93,0)(-0.6,1){6}{\circle*{0.2}}
\put(99,5){\makebox(0,0){$+$}}
\put(102,0){\line(1,0){6}}
\put(105,5){\line(3,5){3}}
\put(114,0){\line(-3,5){3}}
\multiput(102,0)(0.6,1){6}{\circle*{0.2}}
\multiput(108,0)(1,0){7}{\circle*{0.2}}
\multiput(108,0)(0.6,1){6}{\circle*{0.2}}
\multiput(108,0)(-0.6,1){6}{\circle*{0.2}}
\multiput(105,5)(1,0){7}{\circle*{0.2}}
\multiput(111,5)(-0.6,1){6}{\circle*{0.2}}
\end{picture}

\caption{\footnotesize{Illustration for the expression of  $h_2(n+1)$. The multiplication of three on the right-hand-side corresponds to the three possible orientations $SG_2(n+1)$.}} 
\label{hfig}
\end{figure}

\ $\Box$

\bigskip

The values of $f_2(n)$, $g_2(n)$, $h_2(n)$ for small $n$ can be evaluated recursively by Eqs. (\ref{feq}), (\ref{geq}), (\ref{heq}) as listed in Table \ref{tablesg2}. These numbers grow exponentially, and do not have simple integer factorizations, in contrast to the corresponding results for the number of spanning trees \cite{sts}. To estimate the value of the asymptotic growth constant defined in Eq. (\ref{zdef}), we need the following lemmas. For the generalized two-dimensional Sierpinski gasket $SG_{2,b}(n)$, define the ratios
\beq
\alpha_{2,b}(n) = \frac{f_{2,b}(n)}{g_{2,b}(n)} \ , \qquad \beta_{2,b}(n) = \frac{g_{2,b}(n)}{h_{2,b}(n)} \ , 
\label{ratiodef}
\eeq
where $n$ is a non-negative integer. For the ordinary Sierpinski gasket in this section, they are simplified to be $\alpha_2(n)$, $\beta_2(n)$ and their values for small $n$ are listed in Table \ref{tablesg2r}. 

\bigskip

\begin{table}[htbp]
\caption{\label{tablesg2} The first few values of $f_2(n)$, $g_2(n)$, $h_2(n)$.}
\begin{center}
\begin{tabular}{|c||r|r|r|r|}
\hline\hline 
$n$      & 0 &  1 &          2 &                             3 \\ \hline\hline 
$f_2(n)$ & 4 &160 & 13,312,000 & 10,293,452,839,321,600,000,000 \\ \hline 
$g_2(n)$ & 1 & 60 &  7,462,400 &  8,864,355,990,896,640,000,000 \\ \hline 
$h_2(n)$ & 1 & 74 & 13,276,800 & 23,868,720,258,482,176,000,000 \\ \hline\hline 
\end{tabular}
\end{center}
\end{table}

\bigskip

\begin{table}[htbp]
\caption{\label{tablesg2r} The first few values of $\alpha_2(n)$, $\beta_2(n)$. The last digits given are rounded off.}
\begin{center}
\begin{tabular}{|c||r|r|r|r|r|r|}
\hline\hline 
$n$           & 0 &                 1 &                 2 &                    3 &                 4 \\ \hline\hline 
$\alpha_2(n)$ & 4 & 2.66666666666667  & 1.78387650085763  & 1.16121835019855  & 0.736689163182441 \\ \hline 
$\beta_2(n)$  & 1 & 0.810810810810811 & 0.562063147746445 & 0.371379608747416 & 0.238302798822389 \\ \hline\hline 
\end{tabular}
\end{center}
\end{table}

\bigskip

\begin{lemma} \label{lemmasg2ab} For any $n \ge 0$, 
\beq
3\beta_2(n) \le \alpha_2(n) \le 4\beta_2(n) \ .
\eeq
The ratios $\alpha_2(n)$ and $\beta_2(n)$ are both strictly decreasing sequences with the limits
\beq 
\lim _{n \to \infty} \alpha_2(n) = \lim _{n \to \infty} \beta_2(n) = 0 \ .
\eeq
\end{lemma}

{\sl Proof} \quad 
It is clear that $\alpha_2(n)$ and $\beta_2(n)$ cannot be negative. By Eqs. (\ref{feq}) - (\ref{heq}), we have
\beq
\frac{f_2(n+1)}{f_2^2(n)g_2(n)} = 6 + \alpha_2(n) \ ,
\label{feqn}
\eeq
\beq
\frac{g_2(n+1)}{f_2(n)g_2^2(n)} = \frac{\alpha_2(n)}{\beta_2(n)} + 7 + \alpha_2(n) \ ,
\label{geqn}
\eeq
\beq
\frac{h_2(n+1)}{g_2^3(n)} = 12\frac{\alpha_2(n)}{\beta_2(n)} + 14 + 3\alpha_2(n) \ .
\label{heqn}
\eeq
Therefore,
\beq
\alpha_2(n+1) = \frac{\alpha_2(n)\beta_2(n)[6+\alpha_2(n)]} {\alpha_2(n)+7\beta_2(n)+\alpha_2(n)\beta_2(n)} = \alpha_2(n) - \frac{\alpha_2(n)[\alpha_2(n)+\beta_2(n)]} {\alpha_2(n)+7\beta_2(n)+\alpha_2(n)\beta_2(n)} \ ,
\label{alpha}
\eeq
which shows that $\alpha_2(n)$ is strictly decreasing. Similarly, we have
\beq
\beta_2(n+1) = \frac{\alpha_2(n)[\alpha_2(n)+7\beta_2(n)+\alpha_2(n)\beta_2(n)]} {12\alpha_2(n)+14\beta_2(n)+3\alpha_2(n)\beta_2(n)} =
\frac{\alpha_2(n)}{3} - \frac{\alpha_2(n)[3\alpha_2(n)-7\beta_2(n)/3]} {12\alpha_2(n)+14\beta_2(n)+3\alpha_2(n)\beta_2(n)} \ .
\label{beta}
\eeq
With the initial values given in Table \ref{tablesg2r}, $3\beta_2(n) \le \alpha_2(n)$ is proved by induction. By Eqs. (\ref{alpha}) and (\ref{beta}),
\beqs
& & 4\beta_2(n+1) - \alpha_2(n+1) \cr\cr 
& = & \frac{\alpha_2(n)}{3} + \frac{\alpha_2(n)[\alpha_2(n)+\beta_2(n)]} {\alpha_2(n)+7\beta_2(n)+\alpha_2(n)\beta_2(n)} - \frac{\alpha_2(n)[12\alpha_2(n)-28\beta_2(n)/3]} {12\alpha_2(n)+14\beta_2(n)+3\alpha_2(n)\beta_2(n)} \cr\cr
& = & \frac{\alpha_2(n)X(n)} {3[\alpha_2(n)+7\beta_2(n)+\alpha_2(n)\beta_2(n)][12\alpha_2(n)+14\beta_2(n)+3\alpha_2(n)\beta_2(n)]} \ ,
\eeqs
where
\beq
X(n) = 12\alpha_2^2(n) + 48\beta_2(n)[7\beta_2(n)-\alpha_2(n)] +12\alpha_2(n)\beta_2(n)[6\beta_2(n)-\alpha_2(n)] + 3\alpha_2^2(n)\beta_2^2(n) \ge 0
\eeq
such that $\alpha_2(n) \le 4\beta_2(n)$ is proved again by induction. Eq. (\ref{beta}) can be rewritten as
\beq
\beta_2(n+1) = \beta_2(n) - \frac{\alpha_2(n)[4\beta_2(n)-\alpha_2(n)][1+3\beta_2(n)/4]+\alpha_2(n)\beta_2(n)[1-\alpha_2(n)/4]+14\beta_2^2(n)} {12\alpha_2(n)+14\beta_2(n)+3\alpha_2(n)\beta_2(n)} \ ,
\eeq
which shows that $\beta_2(n)$ is strictly decreasing since $\alpha_2(n)$ is less than four, {\textit i.e.} its initial value, for all $n \ge 1$. Eq. (\ref{alpha}) can be rewritten as
\beq
\alpha_2(n+1) = \frac{6\alpha_2(n)}{7} \Big [ 1 - \frac{\alpha_2(n)[1-\beta_2(n)/6]} {\alpha_2(n)+7\beta_2(n)+\alpha_2(n)\beta_2(n)} \Big ] \ ,
\eeq
which is always less than $6\alpha_2(n)/7$ since $\beta_2(n)$ is less than one, {\textit i.e.} its initial value, for all $n \ge 1$ such that $\lim _{n \to \infty} \alpha_2(n)$ is zero. Finally, since $\beta_2(n) \le \alpha_2(n)/3$, $\lim_{n \to \infty} \beta_2(n)$ is zero, and the proof is completed.
\ $\Box$

\bigskip

We notice that the convergences of $\alpha_2(n)$ and $\beta_2(n)$ to zero as $n$ increases are not rapid. The inequality $3\beta_2(n) \le \alpha_2(n)$ can be improved a bit, and we state it as the following lemma.

\bigskip

\begin{lemma} \label{lemmasg2abn} For any $n \ge n_0$, 
\beq
3\beta_2(n) + \frac{\alpha_2^2(n)}{27} \le \alpha_2(n) \ ,
\label{3bla}
\eeq
where $n_0 = inf\{n: \alpha_2(n) \le 3/4, \beta_2(n) \le 1/4 \} =4$.
\end{lemma}

{\sl Proof} \quad 
By Eqs. (\ref{alpha}) and (\ref{beta}), we have
\beqs
& & \alpha_2(n+1) - 3\beta_2(n+1) - \frac{\alpha_2^2(n+1)}{27} \cr\cr
& = & \frac{\alpha_2(n)[9\alpha_2(n)-7\beta_2(n)]} {12\alpha_2(n)+14\beta_2(n)+3\alpha_2(n)\beta_2(n)} - \frac{\alpha_2(n)[\alpha_2(n)+\beta_2(n)]} {\alpha_2(n)+7\beta_2(n)+\alpha_2(n)\beta_2(n)} \cr\cr 
& & - \frac{\alpha_2^2(n)\beta_2^2(n)[6+\alpha_2(n)]^2} {27[\alpha_2(n)+7\beta_2(n)+\alpha_2(n)\beta_2(n)]^2} \cr\cr
& = & \frac{\alpha_2(n)Y(n)} {[\alpha_2(n)+7\beta_2(n)+\alpha_2(n)\beta_2(n)]^2 [12\alpha_2(n)+14\beta_2(n)+3\alpha_2(n)\beta_2(n)]} \ ,
\eeqs
where 
\beqs
Y(n) & = & [\alpha_2(n)+7\beta_2(n)+\alpha_2(n)\beta_2(n)] \Big \{ 3[\alpha_2(n)-3\beta_2(n)][4\beta_2(n)-\alpha_2(n)] \cr\cr 
& & + 3\beta_2(n) \Big [ \alpha_2(n) - 3\beta_2(n) - \frac{\alpha_2^2(n)}{27} \Big ] [3+2\alpha_2(n)] \Big \} \cr\cr
& & + \frac{\alpha_2(n)\beta_2(n)}{27} [ 9\alpha_2^2(n) + 1008\beta_2^2(n) - 153\alpha_2(n)\beta_2(n) + 6\alpha_2^3(n) - 93\alpha_2^2(n)\beta_2(n) \cr\cr 
& & - 60\alpha_2(n)\beta_2^2(n) - 6\alpha_2^3(n)\beta_2(n) - 50\alpha_2^2(n)\beta_2^2(n) - 3\alpha_2^3(n)\beta_2^2(n) ] \cr\cr
& \ge & \frac{\alpha_2(n)\beta_2(n)}{27} \Big \{ 3\alpha_2(n)[\alpha_2(n)-3\beta_2(n)] [3+2\alpha_2(n)] + 252\beta_2(n)[4\beta_2(n)-\alpha_2(n)] \cr\cr 
& & + \alpha_2(n)\beta_2(n) [ 126 - 75\alpha_2(n) - 6\alpha_2^2(n) - 60\beta_2(n) - 50\alpha_2(n)\beta_2(n) - 3\alpha_2^2(n)\beta_2(n) ] \Big \} \ . \cr & &
\eeqs
Because $Y(n)$ is positive whenever $\alpha_2(n) \le 3/4$ and $\beta_2(n) \le 1/4$, which is true for all $n \ge n_0 = 4$ by the previous lemma and Table \ref{tablesg2r}, the inequality is established.
\ $\Box$

\bigskip

We notice that although Eq. (\ref{3bla}) is by no means optimum, it is enough for the following lemma.

\bigskip

\begin{lemma} \label{lemmasg2aob} The sequence of the ratio $\{\alpha_2(n)/\beta_2(n)\}_{n=0}^\infty$ decreases monotonically with the limit
\beq
\lim _{n \to \infty} \alpha_2(n)/\beta_2(n) = 3 \ .
\eeq
\end{lemma}

{\sl Proof} \quad 
The initial value of the ratio is $\alpha_2(0)/\beta_2(0) = 4$. It is clear from Eq. (\ref{beta}) that in the large $n$ limit, the ratio $\alpha_2(n)/\beta_2(n)$ is equal to three. By Eqs. (\ref{alpha}) and (\ref{beta}), we have
\beq
\frac{\alpha_2(n)}{\beta_2(n)} - \frac{\alpha_2(n+1)}{\beta_2(n+1)} = \frac{Z(n)} {\beta_2(n)[\alpha_2(n)+7\beta_2(n)+\alpha_2(n)\beta_2(n)]^2} \ ,
\eeq
where 
\beqs
Z(n) & = & \alpha_2(n)[\alpha_2(n)+7\beta_2(n)+\alpha_2(n)\beta_2(n)]^2 \cr\cr
& & - \beta_2^2(n) [6+\alpha_2(n)] [12\alpha_2(n)+14\beta_2(n)+3\alpha_2(n)\beta_2(n)] \cr\cr
& = & [\alpha_2(n)-3\beta_2(n)] \Big [ 2\alpha_2^2(n)\beta_2(n)+\frac{32}{3}\alpha_2(n)\beta_2^2(n) + \alpha_2^2(n)\beta_2^2(n) \Big ] \cr\cr 
& & + \Big [ \alpha_2(n)-3\beta_2(n)-\frac{\alpha_2^2(n)}{27} \Big ] [\alpha_2^2(n)+17\alpha_2(n)\beta_2(n)+28\beta_2^2(n)] \cr\cr
& & + \frac{\alpha_2^2(n)}{27} [\alpha_2^2(n)+17\alpha_2(n)\beta_2(n)+28\beta_2^2(n)] - \frac{8}{3}\alpha_2^2(n)\beta_2^2(n) \ .
\eeqs
With $\alpha_2(n) \ge 3\beta_2(n)$, $Z(n)$ is positive such that the sequence of the ratio decreases monotonically. 
\ $\Box$

\bigskip

The general expressions for $f_2(n)$ and $g_2(n)$ can be written as follows.

\bigskip

\begin{lemma} \label{lemmasg2fg} For a non-negative integer $m$ and any positive integer $n > m$, 
\beqs
f_2(n) & = & f_2(m)^{\frac{3^{n-m}+1}{2}} g_2(m)^{\frac{3^{n-m}-1}{2}} \prod _{i=1}^{n-m} \Big [ 6 + \alpha_2(n-i) \Big ]^{\frac{3^{i-1}+1}{2}} \cr\cr
& & \times \prod _{j=2}^{n-m} \Big [ 7 + \alpha_2(n-j) + \frac{\alpha_2(n-j)}{\beta_2(n-j)} \Big ]^{\frac{3^{j-1}-1}{2}} \ ,
\label{fgen}
\eeqs
\beqs
g_2(n) & = & f_2(m)^{\frac{3^{n-m}-1}{2}} g_2(m)^{\frac{3^{n-m}+1}{2}} \prod _{i=2}^{n-m} \Big [ 6 + \alpha_2(n-i) \Big ]^{\frac{3^{i-1}-1}{2}} \cr\cr
& & \times \prod _{j=1}^{n-m} \Big [ 7 + \alpha_2(n-j) + \frac{\alpha_2(n-j)}{\beta_2(n-j)} \Big ]^{\frac{3^{j-1}+1}{2}} \ .
\label{ggen}
\eeqs
Here when $n-m=1$, the products with lower limit two are defined to be one.
\end{lemma}

{\sl Proof} \quad 
It is clear from Eqs. (\ref{feqn}) and (\ref{geqn}) that $f_2(m+1)=f_2^2(m)g_2(m)[6+\alpha_2(m)]$ and $g_2(m+1)=f_2(m)g_2^2(m)[7+\alpha_2(m)+\alpha_2(m)/\beta_2(m)]$. Consider Eqs. (\ref{fgen}) and (\ref{ggen}) hold for a certain positive integer $n=k$, then
\beqs
f_2(k+1) & = & f_2^2(k) g_2(k) [6+\alpha_2(k)] \cr\cr
& = & f_2(m)^{3^{k-m}+1} g_2(m)^{3^{k-m}-1} \prod_{i=1}^{k-m} \Big [ 6 + \alpha_2(k-i) \Big ]^{3^{i-1}+1} \cr\cr
& & \times \prod _{j=2}^{k-m} \Big [ 7 + \alpha_2(k-j) + \frac{\alpha_2(k-j)}{\beta_2(k-j)} \Big ]^{3^{j-1}-1} \cr\cr
& & \times f_2(m)^{\frac{3^{k-m}-1}{2}} g_2(m)^{\frac{3^{k-m}+1}{2}} \prod _{i=2}^{k-m} \Big [ 6 + \alpha_2(k-i) \Big ]^{\frac{3^{i-1}-1}{2}} \cr\cr
& & \times \prod _{j=1}^{k-m} \Big [ 7 + \alpha_2(k-j) + \frac{\alpha_2(k-j)}{\beta_2(k-j)} \Big ]^{\frac{3^{j-1}+1}{2}} [6+\alpha_2(k)] \cr\cr
& = & f_2(m)^{\frac{3^{k-m+1}+1}{2}} g_2(m)^{\frac{3^{k-m+1}-1}{2}} [6+\alpha_2(k)] [6+\alpha_2(k-1)]^2 \cr\cr
& & \times \prod_{i=2}^{k-m} \Big [ 6 + \alpha_2(k-i) \Big ]^{\frac{3^i+1}{2}} \Big [ 7 + \alpha_2(k-1) + \frac{\alpha_2(k-1)}{\beta_2(k-1)} \Big ] \cr\cr
& & \times \prod _{j=2}^{k-m} \Big [ 7 + \alpha_2(k-j) + \frac{\alpha_2(k-j)}{\beta_2(k-j)} \Big ]^{\frac{3^j-1}{2}} \cr\cr
& = & f_2(m)^{\frac{3^{k-m+1}+1}{2}} g_2(m)^{\frac{3^{k-m+1}-1}{2}} \prod_{i=1}^{k-m+1} \Big [ 6 + \alpha_2(k+1-i) \Big ]^{\frac{3^{i-1}+1}{2}} \cr\cr
& & \prod _{j=2}^{k-m+1} \Big [ 7 + \alpha_2(k+1-j) + \frac{\alpha_2(k+1-j)}{\beta_2(k+1-j)} \Big ]^{\frac{3^{j-1}-1}{2}} \ ,
\eeqs
and Eq. (\ref{fgen}) is proved by induction. Eq. (\ref{ggen}) can be established by the same procedure.
\ $\Box$

\bigskip

From the above lemmas, we have the following bounds for the asymptotic growth constant.

\bigskip

\begin{lemma} \label{lemmasg2b} The asymptotic growth constant for the number of connected spanning subgraphs on $SG_2(n)$ is bounded:
\beq
\frac{\ln [f_2(m) g_2(m)] + \frac{1}{2} \ln 60} {3^{m+1}} \le z_{SG_2} \le \frac{\ln [f_2(m) g_2(m)] + \frac{1}{2} \ln [6+\alpha_2(m)] \Big [ 7 + \alpha_2(m) + \frac{\alpha_2(m)}{\beta_2(m)} \Big ]} {3^{m+1}} \ ,
\label{zsg2}
\eeq
where $m$ is a positive integer.
\end{lemma}

{\sl Proof} \quad 
By Lemma \ref{lemmasg2fg}, we have
\beq
\ln f_2(n) = \frac{3^{n-m}+1}{2} \ln f_2(m) + \frac{3^{n-m}-1}{2} \ln g_2(m) + \Delta(n,m) \ ,
\eeq
where
\beq
\Delta(n,m) = \sum_{i=1}^{n-m} \frac{3^{i-1}+1}{2} \ln [6+\alpha_2(n-i)] + \sum_{j=2}^{n-m} \frac{3^{j-1}-1}{2} \ln \Big [7 + \alpha_2(n-j) + \frac{\alpha_2(n-j)}{\beta_2(n-j)} \Big ] \ .
\eeq
We have shown that as $m$ increases, $\alpha_2(m)$ decreases to zero in Lemma \ref{lemmasg2ab} and $\alpha_2(m)/\beta_2(m)$ decreases to three in Lemma \ref{lemmasg2aob} such that
\beqs
\Delta(n,m) & \le & \sum_{i=1}^{n-m} \frac{3^{i-1}+1}{2} \ln [6+\alpha_2(m)] + \sum_{j=2}^{n-m} \frac{3^{j-1}-1}{2} \ln \Big [7 + \alpha_2(m) + \frac{\alpha_2(m)}{\beta_2(m)} \Big ] \cr\cr
& = & \frac{1}{2} \Big ( \frac{3^{n-m}-1}{2} +n-m \Big ) \ln [6+\alpha_2(m)] \cr\cr
& & + \frac{1}{2} \Big ( \frac{3^{n-m}-3}{2} -n+m+1 \Big ) \ln \Big [7 + \alpha_2(m) + \frac{\alpha_2(m)}{\beta_2(m)} \Big ] \cr & &
\eeqs
and
\beqs
\Delta(n,m) & \ge & \sum_{i=1}^{n-m} \frac{3^{i-1}+1}{2} \ln 6 + \sum_{j=2}^{n-m} \frac{3^{j-1}-1}{2} \ln 10 \cr\cr
& = & \frac{1}{2} \Big ( \frac{3^{n-m}-1}{2} +n-m \Big ) \ln 6 + \frac{1}{2} \Big ( \frac{3^{n-m}-3}{2} -n+m+1 \Big ) \ln 10 \ .
\eeqs
With the definition for $z_{SG_2}$ given in Eq. (\ref{zdef}) and the number of vertices of $SG_2(n)$ is $3(3^n+1)/2$ by Eq. (\ref{v}), the proof is completed.
\ $\Box$

\bigskip

As $m$ increases, the difference between the upper and lower bounds in Eq. (\ref{zsg2}) becomes small but the convergence is not rapid. We calculate the number of connected spanning subgraphs $f_2(m)$ up to $m=15$, and we have the following proposition.

\bigskip

\begin{propo} \label{proposg2} The asymptotic growth constant for the number of connected spanning subgraphs on the two-dimensional Sierpinski gasket $SG_2(n)$ in the large $n$ limit is $z_{SG_2}=1.276495930...$.

\end{propo}

\bigskip

Without going into details, we state here without proof that the bounds can be improved. For a non-negative integer $m$ and any positive integer $n > m$, the tighter bounds for $\alpha_2(n)$ are
\beq
d^{n-m}(m) \le \alpha_2(n) \le c^{n-m}(m) \ ,
\eeq
where 
\beq
c(m) = \frac{6+\alpha_2(m)}{10+\alpha_2(m)} \ , \qquad 
d(m) = \frac{6\beta_2(m)}{\alpha_2(m)+7\beta_2(m)} \ .
\eeq
It can be shown that 
\beq
z_{SG_2} \le \frac{1}{3^{m+1}} \Big \{ \ln [f_2(m) g_2(m)] + \frac{1}{2} \ln \Big [42 + 6\frac{\alpha_2(m)}{\beta_2(m)} \Big ] + \frac{\alpha_2(m)}{[3-c(m)]} \Big [ \frac{1}{6} + \frac{\beta_2(m)}{7\beta_2(m)+\alpha_2(m)} \Big ] \Big \}
\eeq
and
\beq
z_{SG_2} \ge \frac{1}{3^{m+1}} \Big \{ \ln [f_2(m) g_2(m)] + \frac{\ln 60}{2} + \frac{4\alpha_2(m)}{15[3-d(m)]} - \frac{17\alpha_2^2(m)}{900[3-d^2(m)]} \Big \} \ ,
\eeq
so that the asymptotic growth constant is $z_{SG_2}=1.27649593067...$.

\section{The number of connected spanning subgraphs on $SG_{2,b}(n)$ with $b=3,4$} 
\label{sectionIV}

The method given in the previous section can be applied to the number of connected spanning subgraphs on $SG_{d,b}(n)$ with larger values of $d$ and $b$. The number of configurations to be considered increases as $d$ and $b$ increase, and the recursion relations must be derived individually for each $d$ and $b$. 
In this section, we consider the generalized two-dimensional Sierpinski gasket $SG_{2,b}(n)$ with the number of layers $b$ equal to three and four. 
For $SG_{2,3}(n)$, the numbers of edges and vertices are given by 
\beq
e(SG_{2,3}(n)) = 3 \times 6^n \ ,
\label{esg23}
\eeq
\beq
v(SG_{2,3}(n)) = \frac{7 \times 6^n + 8}{5} \ ,
\label{vsg23}
\eeq
where the three outmost vertices have degree two. There are $(6^n-1)/5$ vertices of $SG_{2,3}(n)$ with degree six and $6(6^n-1)/5$ vertices with degree four. The initial values for the number of connected spanning subgraphs are the same as for $SG_2$: $f_{2,3}(0)=4$, $g_{2,3}(0)=1$ and $h_{2,3}(0)=1$. By the method illustrated in the previous section, we obtain the following recursion relations for any non-negative integer $n$.
\beqs
f_{2,3}(n+1) & = & f_{2,3}^6(n) + 15f_{2,3}^5(n)g_{2,3}(n) + 3f_{2,3}^5(n)h_{2,3}(n) + 78f_{2,3}^4(n)g_{2,3}^2(n) \cr\cr 
& & + 18 f_{2,3}^4(n)g_{2,3}(n)h_{2,3}(n) + 142f_{2,3}^3(n)g_{2,3}^3(n) \ , 
\label{f23eq}
\eeqs
\beqs
g_{2,3}(n+1) & = & f_{2,3}^5(n)g_{2,3}(n) + f_{2,3}^5(n)h_{2,3}(n) + 16f_{2,3}^4(n)g_{2,3}^2(n) + 18f_{2,3}^4(n)g_{2,3}(n)h_{2,3}(n) \cr\cr
& & + 89f_{2,3}^3(n)g_{2,3}^3(n) + 2f_{2,3}^4(n)h_{2,3}^2(n) + 77f_{2,3}^3(n)g_{2,3}^2(n)h_{2,3}(n) \cr\cr 
& & + 171f_{2,3}^2(n)g_{2,3}^4(n) \ , 
\label{g23eq}
\eeqs
\beqs
h_{2,3}(n+1) & = & 3f_{2,3}^4(n)g_{2,3}^2(n) + 6f_{2,3}^4(n)g_{2,3}(n)h_{2,3}(n) + 51f_{2,3}^3(n)g_{2,3}^3(n) + 3f_{2,3}^4(n)h_{2,3}^2(n) \cr\cr & & + 129f_{2,3}^3(n)g_{2,3}^2(n)h_{2,3}(n) + 279f_{2,3}^2(n)g_{2,3}^4(n) + 60f_{2,3}^3(n)g_{2,3}(n)h_{2,3}^2(n) \cr\cr
& & + 564f_{2,3}^2(n)g_{2,3}^3(n)h_{2,3}(n) + 468f_{2,3}(n)g_{2,3}^5(n) \ .
\label{h23eq}
\eeqs
The figures for these configurations are too many to be shown here.
Some values of $f_{2,3}(n)$, $g_{2,3}(n)$, $h_{2,3}(n)$ are listed in Table \ref{tablesg23}. These numbers grow exponentially, and do not have simple integer factorizations.

\bigskip

\begin{table}[htbp]
\caption{\label{tablesg23} The first few values of $f_{2,3}(n)$, $g_{2,3}(n)$, $h_{2,3}(n)$.}
\begin{center}
\begin{tabular}{|c||r|r|r|}
\hline\hline 
$n$          & 0 &      1 &                                      2 \\ \hline\hline 
$f_{2,3}(n)$ & 4 & 56,192 & 1,292,237,078,102,059,106,775,347,494,912 \\ \hline 
$g_{2,3}(n)$ & 1 & 24,624 & 1,015,755,670,321,368,497,188,308,516,864 \\ \hline 
$h_{2,3}(n)$ & 1 & 33,792 & 2,465,934,182,960,517,405,173,530,755,072 \\ \hline\hline 
\end{tabular}
\end{center}
\end{table}

The sequences of the ratios $\{\alpha_{2,3}(n)\}_{n=1}^\infty$ and $\{\beta_{2,3}(n)\}_{n=1}^\infty$ defined in Eq. (\ref{ratiodef}) again decrease monotonically with $\lim _{n \to \infty} \alpha_{2,3}(n) = \beta_{2,3}(n) = 0$. The ratio $\alpha_{2,3}(n)/\beta_{2,3}(n)$ decreases from four to three, the same as the results for $SG_2(n)$. 
The values of $\alpha_{2,3}(n)$, $\beta_{2,3}(n)$ for small $n$ are listed in Table \ref{tablesg23n}. 

\bigskip

\begin{table}[htbp]
\caption{\label{tablesg23n} The first few values of $\alpha_{2,3}(n)$, $\beta_{2,3}(n)$. The last digits given are rounded off.}
\begin{center}
\begin{tabular}{|c||r|r|r|r|r|}
\hline\hline 
$n$               & 0 &                 1 &                 2 &                 3 &                 4 \\ \hline\hline 
$\alpha_{2,3}(n)$ & 4 & 2.28200129954516  & 1.27219282732945  & 0.660858801678112 & 0.326587785819904 \\ \hline 
$\beta_{2,3}(n)$  & 1 & 0.728693181818182 & 0.411915158701392 & 0.215917449918629 & 0.107573237878269 \\ \hline\hline 
\end{tabular}
\end{center}
\end{table}

\bigskip

By the same method as in Lemma \ref{lemmasg2fg}, we have the general expression for the number of connected spanning subgraphs.
\beqs
f_{2,3}(n) & = & f_{2,3}(m)^{\frac{3}{5}(4\times 6^{n-m-1}+1)} g_{2,3}(m)^{\frac{3}{5}(6^{n-m}-1)} \prod_{i=1}^{n-m} P_{2,3}(n-j)^{\frac{3}{5}(4\times 6^{i-2}+1)} \cr\cr
& & \times \prod_{j=2}^{n-m} Q_{2,3}(n-j)^{\frac{3}{5}(6^{j-1}-1)} \ ,
\eeqs
where
\beq
P_{2,3}(m) = 142 + 18\frac{\alpha_{2,3}(m)}{\beta_{2,3}(m)} + 78\alpha_{2,3}(m) + 3\frac{\alpha_{2,3}^2(m)}{\beta_{2,3}(m)} + 15\alpha_{2,3}^2(m) + \alpha_{2,3}^3(m) \ ,
\eeq
\beq
Q_{2,3}(m) = 171 + 77\frac{\alpha_{2,3}(m)}{\beta_{2,3}(m)} + 2\frac{\alpha_{2,3}^2(m)}{\beta_{2,3}^2(m)} + 89\alpha_{2,3}(m) + 18\frac{\alpha_{2,3}^2(m)}{\beta_{2,3}(m)} + 16\alpha_{2,3}^2(m) + \frac{\alpha_{2,3}^3(m)}{\beta_{2,3}(m)} + \alpha_{2,3}^3(m) \ .
\eeq
By the same argument given in Lemma \ref{lemmasg2b}, we have the upper and lower bounds of the asymptotic growth constant for the number of connected spanning subgraphs on $SG_{2,3}(n)$:
\beqs
& & \frac{1}{7\times 6^m} [2\ln f_{2,3}(m) + 3\ln g_{2,3}(m)] + \frac{1}{35\times 6^m} [2\ln 196 + 3\ln 420] \le z_{SG_{2,3}} \cr\cr
& & \le \frac{1}{7\times 6^m} [2\ln f_{2,3}(m) + 3\ln g_{2,3}(m)] + \frac{1}{35\times 6^m} [2\ln P_{2,3}(m) + 3\ln Q_{2,3}(m)]  \ ,
\label{zsg23}
\eeqs
with $m$ a positive integer. We have the following proposition.

\bigskip

\begin{propo} \label{proposg23} The asymptotic growth constant for the number of connected spanning subgraphs on the two-dimensional Sierpinski gasket $SG_{2,3}(n)$ in the large $n$ limit is $z_{SG_{2,3}}=1.3972789680...$.

\end{propo}

\bigskip

\noindent We notice that the convergence of the upper and lower bounds remains slow. 

\bigskip

For $SG_{2,4}(n)$, the numbers of edges and vertices are given by 
\beq
e(SG_{2,4}(n)) = 3 \times 10^n \ ,
\label{esg24}
\eeq
\beq
v(SG_{2,4}(n)) = \frac{4 \times 10^n + 5}{3} \ ,
\label{vsg24}
\eeq
where again the three outmost vertices have degree two. There are $(10^n-1)/3$ vertices of $SG_{2,4}(n)$ with degree six, and $(10^n-1)$ vertices with degree four. The initial values for the number of connected spanning subgraphs are the same as for $SG_2$: $f_{2,4}(0)=4$, $g_{2,4}(0)=1$ and $h_{2,4}(0)=1$.
We wrote a computer program to obtain the recursion relations for $SG_{2,4}(n)$. They are lengthy and given in the appendix. Some values of $f_{2,4}(n)$, $g_{2,4}(n)$, $h_{2,4}(n)$ are listed in Table \ref{tablesg24}. These numbers grow exponentially, and do not have simple integer factorizations.

\bigskip

\begin{table}[htbp]
\caption{\label{tablesg24} The first few values of $f_{2,4}(n)$, $g_{2,4}(n)$, $h_{2,4}(n)$.}
\begin{center}
\begin{tabular}{|c||r|r|}
\hline\hline 
$n$          & 1 & 2 \\ \hline \hline
$f_{2,4}(n)$ & 164,119,040 & \tiny 27,140,375,625,882,898,681,725,275,604,427,985,839,201,951,967,246,962,831,668,354,270,630,763,784,348,631,040,000 \\ \hline 
$g_{2,4}(n)$ &  77,622,016 & \tiny 25,675,411,803,142,714,297,950,351,525,972,498,833,548,895,007,181,465,816,231,861,389,426,797,930,493,247,488,000 \\ \hline 
$h_{2,4}(n)$ & 112,848,672 & \tiny 74,273,341,808,825,211,957,637,724,253,224,196,638,029,720,486,058,269,503,940,976,372,670,504,798,196,334,592,000 \\ \hline\hline 
\end{tabular}
\end{center}
\end{table}

\bigskip

The sequences of the ratios $\{\alpha_{2,4}(n)\}_{n=1}^\infty$ and $\{\beta_{2,4}(n)\}_{n=1}^\infty$ defined in Eq. (\ref{ratiodef}) again decrease monotonically with $\lim _{n \to \infty} \alpha_{2,4}(n) = \beta_{2,4}(n) = 0$. The ratio $\alpha_{2,4}(n)/\beta_{2,4}(n)$ decreases from four to three. The values of $\alpha_{2,4}(n)$, $\beta_{2,4}(n)$ for small $n$ are listed in Table \ref{tablesg24n}. 

\bigskip

\begin{table}[htbp]
\caption{\label{tablesg24n} The first few values of $\alpha_{2,4}(n)$, $\beta_{2,4}(n)$. The last digits given are rounded off.}
\begin{center}
\begin{tabular}{|c||r|r|r|r|r|}
\hline\hline 
$n$               & 0 &                 1 &                 2 &                 3 &                 4 \\ \hline\hline 
$\alpha_{2,4}(n)$ & 4 & 2.11433622131123  & 1.05705707211134  & 0.475214294459902 & 0.199993476915309 \\ \hline 
$\beta_{2,4}(n)$  & 1 & 0.687841643364664 & 0.345688118749653 & 0.156412715166630 & 0.662275943767262 \\ \hline\hline 
\end{tabular}
\end{center}
\end{table}

\bigskip

By the same method as in Lemma \ref{lemmasg2fg}, we have the general expression for the number of connected spanning subgraphs.
\beq
f_{2,4}(n) = f_{2,4}(m)^{\frac{1}{3}(10^{n-m}+2)} g_{2,4}(m)^{\frac{2}{3}(10^{n-m}-1)} \prod_{i=1}^{n-m} P_{2,4}(n-j)^{\frac{1}{3}(10^{i-1}+2)} \prod_{j=2}^{n-m} Q_{2,4}(n-j)^{\frac{2}{3}(10^{j-1}-1)} \ ,
\eeq
where
\beqs
P_{2,4}(m) & = & 11354 + 5856\frac{\alpha_{2,4}(m)}{\beta_{2,4}(m)} + 516\frac{\alpha_{2,4}^2(m)}{\beta_{2,4}^2(m)} + 2\frac{\alpha_{2,4}^3(m)}{\beta_{2,4}^3(m)} + 13626\alpha_{2,4}(m) + 4140\frac{\alpha_{2,4}^2(m)}{\beta_{2,4}(m)} \cr\cr 
& & + 174\frac{\alpha_{2,4}^3(m)}{\beta_{2,4}^2(m)} + 6936\alpha_{2,4}^2(m) + 1140\frac{\alpha_{2,4}^3(m)}{\beta_{2,4}(m)} + 15\frac{\alpha_{2,4}^4(m)}{\beta_{2,4}^2(m)} + 1928\alpha_{2,4}^3(m) \cr\cr 
& & + 144\frac{\alpha_{2,4}^4(m)}{\beta_{2,4}(m)} + 309\alpha_{2,4}^4(m) + 7\frac{\alpha_{2,4}^5(m)}{\beta_{2,4}(m)} + 27\alpha_{2,4}^5(m) + \alpha_{2,4}^6(m) \ ,
\eeqs
\beqs
Q_{2,4}(m) & = & 13732 + 14480\frac{\alpha_{2,4}(m)}{\beta_{2,4}(m)} + 2786\frac{\alpha_{2,4}^2(m)}{\beta_{2,4}^2(m)} + 82\frac{\alpha_{2,4}^3(m)}{\beta_{2,4}^3(m)} + 16250\alpha_{2,4}(m) \cr\cr 
& & + 10609\frac{\alpha_{2,4}^2(m)}{\beta_{2,4}(m)} + 1095\frac{\alpha_{2,4}^3(m)}{\beta_{2,4}^2(m)} + 12\frac{\alpha_{2,4}^4(m)}{\beta_{2,4}^3(m)} + 8015\alpha_{2,4}^2(m) + 3130\frac{\alpha_{2,4}^3(m)}{\beta_{2,4}(m)} \cr\cr 
& & + 142\frac{\alpha_{2,4}^4(m)}{\beta_{2,4}^2(m)} + 2148\alpha_{2,4}^3(m) + 462\frac{\alpha_{2,4}^4(m)}{\beta_{2,4}(m)} + 6\frac{\alpha_{2,4}^5(m)}{\beta_{2,4}^2(m)} + 332\alpha_{2,4}^4(m) \cr\cr 
& & + 34\frac{\alpha_{2,4}^5(m)}{\beta_{2,4}(m)} + 28\alpha_{2,4}^5(m) + \frac{\alpha_{2,4}^6(m)}{\beta_{2,4}(m)} + \alpha_{2,4}^6(m)  \ .
\eeqs
By the same argument given in Lemma \ref{lemmasg2b}, we have the upper and lower bounds of the asymptotic growth constant for the number of connected spanning subgraphs on $SG_{2,4}(n)$:
\beqs
& & \frac{1}{4\times 10^m} [\ln f_{2,4}(m) + 2\ln g_{2,4}(m)] + \frac{1}{36\times 10^m} [\ln 33620 + 2\ln 84460] \le z_{SG_{2,4}} \cr\cr
& & \le \frac{1}{4\times 10^m} [\ln f_{2,4}(m) + 2\ln g_{2,4}(m)] + \frac{1}{36\times 10^m} [\ln P_{2,4}(m) + 2\ln Q_{2,4}(m)]  \ ,
\label{zsg24}
\eeqs
with $m$ a positive integer. We have the following proposition.

\bigskip

\begin{propo} \label{proposg24} The asymptotic growth constant for the number of connected spanning subgraphs on the two-dimensional Sierpinski gasket $SG_{2,4}(n)$ in the large $n$ limit is $z_{SG_{2,4}}=1.4849112...$.

\end{propo}

\bigskip

\noindent Here the convergence of the upper and lower bounds is again slow. 

\section{The number of connected spanning subgraphs on $SG_d(n)$ with $d=3,4$} 
\label{sectionV}

In this section, we derive the asymptotic growth constant of connected spanning subgraphs on $SG_d(n)$ with $d=3,4$.
For the three-dimensional Sierpinski gasket $SG_3(n)$, we use the following definitions.

\bigskip

\begin{defi} \label{defisg3} Consider the three-dimensional Sierpinski gasket $SG_3(n)$ at stage $n$. (i) Define $f_3(n) \equiv N_{CSSG}(SG_3(n))$ as the number of connected spanning subgraphs. (ii) Define $g_3(n)$ as the number of spanning subgraphs with two components such that one certain outmost vertices belongs to one component and the other three outmost vertices belong to another component. (iii) Define $h_3(n)$ as the number of spanning subgraphs with two components such that two certain outmost vertices belong to one component and the other two outmost vertices belong to another component. (iv) Define $r_3(n)$ as the number of spanning subgraphs with three components such that two certain outmost vertices belong to one component and the other two outmost vertices separately belong to other components. (v) Define $s_3(n)$ as the number of spanning subgraphs with four components such that each of the outmost vertices belongs to a different component.
\end{defi}

\bigskip

\begin{figure}[htbp]
\unitlength 1.8mm 
\begin{picture}(66,5)
\put(0,0){\line(1,0){6}}
\put(0,0){\line(3,5){3}}
\put(6,0){\line(-3,5){3}}
\put(0,0){\line(3,2){3}}
\put(6,0){\line(-3,2){3}}
\put(3,2){\line(0,1){3}}
\put(3,-2){\makebox(0,0){$f_3(n)$}}
\put(12,0){\line(1,0){6}}
\multiput(12,0)(0.3,0.5){11}{\circle*{0.2}}
\multiput(18,0)(-0.3,0.5){11}{\circle*{0.2}}
\multiput(15,2)(0,0.5){7}{\circle*{0.2}}
\put(12,0){\line(3,2){3}}
\put(18,0){\line(-3,2){3}}
\put(15,-2){\makebox(0,0){$g_3(n)$}}
\put(24,0){\line(1,0){6}}
\multiput(24,0)(0.3,0.5){11}{\circle*{0.2}}
\multiput(30,0)(-0.3,0.5){11}{\circle*{0.2}}
\put(27,2){\line(0,1){3}}
\multiput(24,0)(0.6,0.4){6}{\circle*{0.2}}
\multiput(30,0)(-0.6,0.4){6}{\circle*{0.2}}
\put(27,-2){\makebox(0,0){$h_3(n)$}}
\put(36,0){\line(1,0){6}}
\multiput(36,0)(0.3,0.5){11}{\circle*{0.2}}
\multiput(42,0)(-0.3,0.5){11}{\circle*{0.2}}
\multiput(39,2)(0,0.5){7}{\circle*{0.2}}
\multiput(36,0)(0.6,0.4){6}{\circle*{0.2}}
\multiput(42,0)(-0.6,0.4){6}{\circle*{0.2}}
\put(39,-2){\makebox(0,0){$r_3(n)$}}
\multiput(48,0)(0.5,0){13}{\circle*{0.2}}
\multiput(48,0)(0.3,0.5){11}{\circle*{0.2}}
\multiput(54,0)(-0.3,0.5){11}{\circle*{0.2}}
\multiput(51,2)(0,0.5){7}{\circle*{0.2}}
\multiput(48,0)(0.6,0.4){6}{\circle*{0.2}}
\multiput(54,0)(-0.6,0.4){6}{\circle*{0.2}}
\put(51,-2){\makebox(0,0){$s_3(n)$}}
\end{picture}

\vspace*{5mm}
\caption{\footnotesize{Illustration for the spanning subgraphs $f_3(n)$, $g_3(n)$, $h_3(n)$, $r_3(n)$ and $s_3(n)$. The two outmost vertices at the ends of a solid line belong to one component, while the two outmost vertices at the ends of a dot line belong to separated components.}} 
\label{fghrsfig}
\end{figure}
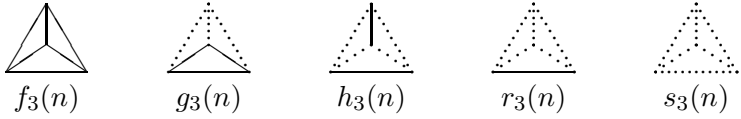

The quantities $f_3(n)$, $g_3(n)$, $h_3(n)$, $r_3(n)$ and $s_3(n)$ are illustrated in Fig. \ref{fghrsfig}, where only the outmost vertices are shown. There are four different classes of connected subgraphs enumerated by $g_3(n)$, three classes enumerated by $h_3(n)$, and six classes enumerated by $r_3(n)$. 
The initial values at stage zero are $f_3(0)=38$, $g_3(0)=4$, $h_3(0)=1$, $r_3(0)=1$ and $s_3(0)=1$. 

\bigskip

\bigskip

The recursion relations are lengthy and given in the appendix. Some values of $f_3(n)$, $g_3(n)$, $h_3(n)$, $r_3(n)$, $s_3(n)$ are listed in Table \ref{tablesg3}. These numbers grow exponentially, and do not have simple integer factorizations.

\bigskip

\begin{table}[htbp]
\caption{\label{tablesg3} The first few values of $f_3(n)$, $g_3(n)$, $h_3(n)$, $r_3(n)$, $s_3(n)$.}
\begin{center}
\begin{tabular}{|c||r|r|r|}
\hline\hline 
$n$      & 0 &      1 &                                      2 \\ \hline\hline 
$f_3(n)$ & 38 & 8,554,560 & 25,988,410,915,610,195,960,527,441,920 \\ \hline 
$g_3(n)$ &  4 & 1,271,416 &  4,544,490,996,892,396,578,747,598,336 \\ \hline 
$h_3(n)$ &  1 &    39,502 &     73,629,059,909,730,939,289,401,696 \\ \hline
$r_3(n)$ &  1 &   254,462 &    917,115,147,969,863,922,701,973,216 \\ \hline 
$s_3(n)$ &  1 &   153,824 &    637,427,406,318,067,141,227,862,784 \\ \hline\hline 
\end{tabular}
\end{center}
\end{table}

\bigskip

We find it is difficult to derive the bounds of the asymptotic growth constant for the number of connected spanning subgraphs on $SG_3(n)$. We calculate $f_3(m)$ up to $m=10$, and fit the numerical value of the asymptotic growth constant to have the following proposition.

\bigskip

\begin{propo} \label{proposg3} The asymptotic growth constant for the number of connected spanning subgraphs on the three-dimensional Sierpinski gasket $SG_3(n)$ in the large $n$ limit is $z_{SG_3}=2.06371$.

\end{propo}

\bigskip

For the four-dimensional Sierpinski gasket $SG_4(n)$, we use the following definitions.

\bigskip

\begin{defi} \label{defisg4} Consider the four-dimensional Sierpinski gasket $SG_4(n)$ at stage $n$. (i) Define $f_4(n) \equiv N_{CSSG}(SG_4(n))$ as the number of connected spanning subgraphs. (ii) Define $g_4(n)$ as the number of spanning subgraphs with two components such that two certain outmost vertices belong to one component and the other three outmost vertices belong to another component. (iii) Define $g^\prime_4(n)$ as the number of spanning subgraphs with two components such that one certain outmost vertices belong to one component and the other four outmost vertices belong to another component. (iv) Define $h_4(n)$ as the number of spanning subgraphs with three components such that one certain outmost vertices belong to one component, two certain other outmost vertices belong to another component and the remaining two outmost vertices belong to a third component. (v) Define $h^\prime_4(n)$ as the number of spanning subgraphs with three components such that three certain outmost vertices belong to one component and the other two outmost vertices separately belong to other components. (vi) Define $r_4(n)$ as the number of spanning subgraphs with four components such that two certain outmost vertices belong to one component and the other three outmost vertices separately belong to other components. (vii) Define $s_4(n)$ as the number of spanning subgraphs with five components such that each of the outmost vertices belongs to a different component.
\end{defi}

\bigskip

The quantities $f_4(n)$, $g_4(n)$, $g^\prime_4(n)$, $h_4(n)$, $h^\prime_4(n)$, $r_4(n)$ and $s_4(n)$ are illustrated in Fig. \ref{fghrsfign}, where only the outmost vertices are shown. There are ten different classes of connected subgraphs enumerated by $g_4(n)$, five classes enumerated by $g^\prime_4(n)$, fifteen classes enumerated by $h_4(n)$, ten classes enumerated by $h^\prime_4(n)$ and ten classes enumerated by $r_4(n)$. 
The initial values at stage zero are $f_4(0)=728$, $g_4(0)=4$, $g^\prime_4(0)=38$, $h_4(0)=1$, $h^\prime_4(0)=4$, $r_4(0)=1$ and $s_4(0)=1$.

\bigskip

\begin{figure}[htbp]
\unitlength 1.8mm 
\begin{picture}(58,9)
\put(2,0){\line(1,0){6}}
\put(2,0){\line(4,3){8}}
\put(2,0){\line(-1,3){2}}
\put(2,0){\line(1,3){3}}
\put(8,0){\line(-1,3){3}}
\put(8,0){\line(1,3){2}}
\put(8,0){\line(-4,3){8}}
\put(0,6){\line(1,0){10}}
\put(5,9){\line(5,-3){5}}
\put(5,9){\line(-5,-3){5}}
\put(5,-2){\makebox(0,0){$f_4(n)$}}
\put(18,0){\line(1,0){6}}
\multiput(18,0)(0.4,0.3){21}{\circle*{0.2}}
\multiput(18,0)(-0.2,0.6){11}{\circle*{0.2}}
\multiput(18,0)(0.2,0.6){16}{\circle*{0.2}}
\multiput(24,0)(-0.2,0.6){16}{\circle*{0.2}}
\multiput(24,0)(0.2,0.6){11}{\circle*{0.2}}
\multiput(24,0)(-0.4,0.3){21}{\circle*{0.2}}
\put(16,6){\line(1,0){10}}
\put(21,9){\line(5,-3){5}}
\put(21,9){\line(-5,-3){5}}
\put(21,-2){\makebox(0,0){$g_4(n)$}}
\put(34,0){\line(1,0){6}}
\put(34,0){\line(4,3){8}}
\put(34,0){\line(-1,3){2}}
\multiput(34,0)(0.2,0.6){16}{\circle*{0.2}}
\multiput(40,0)(-0.2,0.6){16}{\circle*{0.2}}
\put(40,0){\line(1,3){2}}
\put(40,0){\line(-4,3){8}}
\put(32,6){\line(1,0){10}}
\multiput(37,9)(0.5,-0.3){11}{\circle*{0.2}}
\multiput(37,9)(-0.5,-0.3){11}{\circle*{0.2}}
\put(37,-2){\makebox(0,0){$g^\prime_4(n)$}}
\put(50,0){\line(1,0){6}}
\multiput(50,0)(0.4,0.3){21}{\circle*{0.2}}
\multiput(50,0)(-0.2,0.6){11}{\circle*{0.2}}
\multiput(50,0)(0.2,0.6){16}{\circle*{0.2}}
\multiput(56,0)(-0.2,0.6){16}{\circle*{0.2}}
\multiput(56,0)(0.2,0.6){11}{\circle*{0.2}}
\multiput(56,0)(-0.4,0.3){21}{\circle*{0.2}}
\put(48,6){\line(1,0){10}}
\multiput(53,9)(0.5,-0.3){11}{\circle*{0.2}}
\multiput(53,9)(-0.5,-0.3){11}{\circle*{0.2}}
\put(53,-2){\makebox(0,0){$h_4(n)$}}
\end{picture}

\vspace*{10mm}

\begin{picture}(58,9)
\multiput(2,0)(0.5,0){13}{\circle*{0.2}}
\multiput(2,0)(0.4,0.3){21}{\circle*{0.2}}
\multiput(2,0)(-0.2,0.6){11}{\circle*{0.2}}
\multiput(2,0)(0.2,0.6){16}{\circle*{0.2}}
\multiput(8,0)(-0.2,0.6){16}{\circle*{0.2}}
\multiput(8,0)(0.2,0.6){11}{\circle*{0.2}}
\multiput(8,0)(-0.4,0.3){21}{\circle*{0.2}}
\put(0,6){\line(1,0){10}}
\put(5,9){\line(5,-3){5}}
\put(5,9){\line(-5,-3){5}}
\put(5,-2){\makebox(0,0){$h^\prime_4(n)$}}
\put(18,0){\line(1,0){6}}
\multiput(18,0)(0.4,0.3){21}{\circle*{0.2}}
\multiput(18,0)(-0.2,0.6){11}{\circle*{0.2}}
\multiput(18,0)(0.2,0.6){16}{\circle*{0.2}}
\multiput(24,0)(-0.2,0.6){16}{\circle*{0.2}}
\multiput(24,0)(0.2,0.6){11}{\circle*{0.2}}
\multiput(24,0)(-0.4,0.3){21}{\circle*{0.2}}
\multiput(16,6)(0.5,0){21}{\circle*{0.2}}
\multiput(21,9)(0.5,-0.3){11}{\circle*{0.2}}
\multiput(21,9)(-0.5,-0.3){11}{\circle*{0.2}}
\put(21,-2){\makebox(0,0){$r_4(n)$}}
\multiput(34,0)(0.5,0){13}{\circle*{0.2}}
\multiput(34,0)(0.4,0.3){21}{\circle*{0.2}}
\multiput(34,0)(-0.2,0.6){11}{\circle*{0.2}}
\multiput(34,0)(0.2,0.6){16}{\circle*{0.2}}
\multiput(40,0)(-0.2,0.6){16}{\circle*{0.2}}
\multiput(40,0)(0.2,0.6){11}{\circle*{0.2}}
\multiput(40,0)(-0.4,0.3){21}{\circle*{0.2}}
\multiput(32,6)(0.5,0){21}{\circle*{0.2}}
\multiput(37,9)(0.5,-0.3){11}{\circle*{0.2}}
\multiput(37,9)(-0.5,-0.3){11}{\circle*{0.2}}
\put(37,-2){\makebox(0,0){$s_4(n)$}}
\end{picture}

\vspace*{5mm}
\caption{\footnotesize{Illustration for the spanning subgraphs $f_4(n)$, $g_4(n)$, $g^\prime_4(n)$, $h_4(n)$, $h^\prime_4(n)$, $r_4(n)$ and $s_4(n)$. The two outmost vertices at the ends of a solid line belong to one component, while the two outmost vertices at the ends of a dot line belong to separated components.}} 
\label{fghrsfign}
\end{figure}

\bigskip

We wrote a computer program to obtain the recursion relations for $SG_4(n)$. They are too lengthy to be included here, and are available from the authors on request. Some values of $f_4(n)$, $g_4(n)$, $g^\prime_4(n)$, $h_4(n)$, $h^\prime_4(n)$, $r_4(n)$, $s_4(n)$ are listed in Table \ref{tablesg4}. These numbers grow exponentially, and do not have simple integer factorizations.

\bigskip

\begin{table}[htbp]
\caption{\label{tablesg4} The first few values of $f_4(n)$, $g_4(n)$, $g^\prime_4(n)$, $h_4(n)$, $h^\prime_4(n)$, $r_4(n)$, $s_4(n)$.}
\begin{center}
\begin{tabular}{|c||r|r|}
\hline\hline 
$n$             &              1 & 2 \\ \hline \hline
$f_4(n)$  & 778,626,762,895,872 & \tiny 1,024,406,418,765,003,907,906,096,145,114,250,200,136,082,865,744,856,739,402,552,777,712,697,606,144 \\ \hline 
$g_4(n)$        &     88,489,486,528 & \tiny 3,197,766,124,028,071,576,597,031,293,816,293,624,011,891,902,039,451,422,566,691,192,700,928 \\ \hline 
$g^\prime_4(n)$ & 52,683,007,497,792 & \tiny 69,863,645,008,967,428,965,504,302,095,638,435,727,081,373,061,446,549,243,263,724,682,869,735,424 \\ \hline
$h_4(n)$        &     15,626,116,736 & \tiny 482,200,982,250,980,661,780,613,757,386,400,225,524,299,614,671,798,975,981,818,835,107,840 \\ \hline 
$h^\prime_4(n)$ &  3,629,303,504,832 & \tiny 4,765,691,494,696,738,414,627,738,223,389,422,884,987,040,021,355,690,863,179,839,765,575,892,992 \\ \hline 
$r_4(n)$        &    258,767,297,696 & \tiny 325,229,810,040,355,155,302,761,176,752,191,820,409,762,202,792,611,492,002,077,131,210,227,712 \\ \hline 
$s_4(n)$        &     94,459,269,024 & \tiny 110,974,534,976,153,854,286,043,758,382,592,092,762,465,813,295,695,669,459,295,951,908,765,696 \\ 
\hline\hline 
\end{tabular}
\end{center}
\end{table}

\bigskip

It is even more difficult to derive the bounds of the asymptotic growth constant for the number of connected spanning subgraphs on $SG_4(n)$. We calculate $f_4(m)$ up to $m=6$, and we are satisfied with numerical fitting of the asymptotic growth constant to have the following proposition.

\bigskip

\begin{propo} \label{proposg4} The asymptotic growth constant for the number of connected spanning subgraphs on the four-dimensional Sierpinski gasket $SG_4(n)$ in the large $n$ limit is $z_{SG_4}=2.7686$.

\end{propo}

\bigskip

\section{Discussion}

Compared with the asymptotic growth constant for the number of spanning forests $N_{SF}$ on the Sierpinski gasket in Ref. \cite{sfs}, we find that $N_{CSSG}$ is larger than $N_{SF}$ for all the considered cases. We conjecture that this inequality holds for all the generalized Sierpinski gasket. Define
\beq
\tilde z_G = \lim_{v(G) \to \infty} \frac{\ln N_{SF}(G)}{v(G)} \ .
\label{zdefsf}
\eeq
We list the first few values of $\tilde z_{SG_d}$, $z_{SG_d}$, and their ratios in Table \ref{zsgdtable}.

As the spanning tree is a special case of connected spanning subgraphs where there is no cycles allowed, the number of spanning trees $N_{ST}(G)$ is always less than $N_{CSSG}(G)$. Define
\beq
\underline z_G = \lim_{v(G) \to \infty} \frac{\ln N_{ST}(G)}{v(G)} \ ,
\label{zdefn}
\eeq
then $\underline z_G < z_G$. We have obtained such asymptotic growth constants for the number of spanning trees on the Sierpinski gasket $SG_d$ for general $d$ and $SG_{2,b}$ with $b=3,4$ in Ref. \cite{sts}. They serve as the lower bounds for our current consideration for the connected spanning subgraphs.
We list the first few values of $\underline z_{SG_d}$, $z_{SG_d}$, and their ratios in Table \ref{zsgdtable}. Notice that lower bound $\underline z_{SG_d}$ is closer to the exact value $z_{SG_d}$ when $d$ is small, in contrast to the results for the spanning forests given in \cite{sfs} that $\underline z_{SG_d}$ is closer to $\tilde z_G$ when $d$ is large.

\bigskip

\begin{table}
\caption{\label{zsgdtable} Numerical values of $\underline z_{SG_d}$, $\tilde z_G$, $z_{SG_d}$, and their ratios. The last digits given are rounded off.}
\begin{center}
\begin{tabular}{|c|c|c|c|c|c|c|}
\hline\hline 
$d$ &  $D$  & $\underline z_{SG_d}$ & $\tilde z_G$ & $z_{SG_d}$ & $\underline z_{SG_d}/z_{SG_d}$ & $\tilde z_{SG_d}/z_{SG_d}$ \\ \hline\hline 
2   & 1.585 & 1.048594857 & 1.247337199 & 1.276495931 & 0.8214635326 & 0.9771572077 \\ \hline
3   & 2     & 1.569396409 & 1.666806281 & 2.06371     & 0.760475     & 0.807675 \\ \hline
4   & 2.322 & 1.914853265 & 1.981017076 & 2.7686      & 0.69163      & 0.71553 \\ \hline\hline 
\end{tabular}
\end{center}
\end{table}
\bigskip

Acknowledgements:
The research of S.C.C. was partially supported by the NSC grant NSC-96-2112-M-006-001. The research of L.C.C was partially supported by TJ \& MY Foundation and the NSC grant NSC-96-2115-M-030-002. L.C.C. would like to thank PIMS, university of British Columbia for the hospitality.

\bigskip

\appendix

\section{Recursion relations for $SG_{2,4}(n)$}

We give the recursion relations for the generalized two-dimensional Sierpinski gasket $SG_{2,4}(n)$ here. Since the subscript is $(d,b)=(2,4)$ for all the quantities throughout this section, we will use the simplified notation $f_{n+1}$ to denote $f_{2,4}(n+1)$ and similar notations for other quantities. For any non-negative integer $n$, we have
\beqs
f_{n+1} & = & f_n^{10} + 27f_n^9g_n + 7f_n^9h_n + 309f_n^8g_n^2 + 144f_n^8g_nh_n + 1928f_n^7g_n^3 + 15f_n^8h_n^2 + 1140f_n^7g_n^2h_n \cr\cr
& & + 6936f_n^6g_n^4 + 174f_n^7g_nh_n^2 + 4140f_n^6g_n^3h_n + 13626f_n^5g_n^5 + 2f_n^7h_n^3 + 516f_n^6g_n^2h_n^2 \cr\cr
& & + 5856f_n^5g_n^4h_n + 11354f_n^4g_n^6 \ ,
\label{f24eq}
\eeqs
\beqs
g_{n+1} & = & f_n^9g_n + f_n^9h_n + 28f_n^8g_n^2 + 34f_n^8g_nh_n + 332f_n^7g_n^3 + 6f_n^8h_n^2 + 462f_n^7g_n^2h_n + 2148f_n^6g_n^4 \cr\cr
& & + 142f_n^7g_nh_n^2 + 3130f_n^6g_n^3h_n + 8015f_n^5g_n^5 + 12f_n^7h_n^3 + 1095f_n^6g_n^2h_n^2 + 10609f_n^5g_n^4h_n \cr\cr
& & + 16250f_n^4g_n^6 + 82f_n^6g_nh_n^3 + 2786f_n^5g_n^3h_n^2 + 14480f_n^4g_n^5h_n + 13732f_n^3g_n^7 \ ,
\label{g24eq}
\eeqs
\beqs
h_{n+1} & = & 3f_n^8g_n^2 + 6f_n^8g_nh_n + 87f_n^7g_n^3 + 3f_n^8h_n^2 + 189f_n^7g_n^2h_n + 1068f_n^6g_n^4 + 117f_n^7g_nh_n^2 \cr\cr
& & + 2558f_n^6g_n^3h_n + 7113f_n^5g_n^5 + 15f_n^7h_n^3 + 1869f_n^6g_n^2h_n^2 + 17763f_n^5g_n^4h_n + 26934f_n^4g_n^6 \cr\cr
& & + 444f_n^6g_nh_n^3 + 12756f_n^5g_n^3h_n^2 + 61422f_n^4g_n^5h_n + 53826f_n^3g_n^7 + 20f_n^6h_n^4 + 2388f_n^5g_n^2h_n^3 \cr\cr
& & + 30948f_n^4g_n^4h_n^2 + 83234f_n^3g_n^6h_n + 42210f_n^2g_n^8 \ .
\label{h24eq}
\eeqs

\section{Recursion relations for $SG_3(n)$}

We give the recursion relations for the three-dimensional Sierpinski gasket $SG_3(n)$ here. Since the subscript is $d=3$ for all the quantities throughout this section, we will use the simplified notation $f_{n+1}$ to denote $f_3(n+1)$ and similar notations for other quantities. For any non-negative integer $n$, we have
\beqs
f_{n+1} & = & f_n^4 + 12f_n^3g_n + 12f_n^3h_n + 12f_n^3r_n + 48f_n^2g_n^2 + 96f_n^2g_nh_n + 48f_n^2h_n^2 + 72f_n^2g_nr_n \cr\cr
& & + 72f_n^2h_nr_n + 56f_ng_n^3 + 168f_ng_n^2h_n + 168f_ng_nh_n^2 + 56f_nh_n^3 \ , 
\label{f3eq}
\eeqs
\beqs
g_{n+1} & = & f_n^3g_n + 3f_n^3r_n + 9f_n^2g_n^2 + 12f_n^2g_nh_n + f_n^3s_n + 36f_n^2g_nr_n + 30f_n^2h_nr_n + 28f_ng_n^3 \cr\cr
& & + 66f_ng_n^2h_n + 54f_ng_nh_n^2 + 6f_n^2g_ns_n + 6f_n^2h_ns_n + 24f_n^2r_n^2 + 108f_ng_n^2r_n \cr\cr
& & + 192f_ng_nh_nr_n + 84f_nh_n^2r_n + 20g_n^4 + 72g_n^3h_n + 96g_n^2h_n^2 + 56g_nh_n^3 \ ,
\label{g3eq}
\eeqs
\beqs
h_{n+1} & = & 2f_n^2h_n^2 + 4f_n^2h_nr_n + 12f_ng_n^2h_n + 12f_ng_nh_n^2 + 16f_nh_n^3 + 2f_n^2r_n^2 + 12f_ng_n^2r_n \cr\cr
& & + 48f_ng_nh_nr_n + 36f_nh_n^2r_n + 2g_n^4 + 16g_n^3h_n + 36g_n^2h_n^2 + 32g_nh_n^3 + 22h_n^4 \ , 
\label{h3eq}
\eeqs
\beqs
r_{n+1} & = & f_n^2g_n^2 + 2f_n^2h_n^2 + 6f_n^2g_nr_n + 6f_n^2h_nr_n + 6f_ng_n^3 + 22f_ng_n^2h_n + 14f_ng_nh_n^2 + 16f_nh_n^3 \cr\cr
& & + 2f_n^2g_ns_n +2f_n^2h_ns_n + 12f_n^2r_n^2 + 60f_ng_n^2r_n + 132f_ng_nh_nr_n + 66f_nh_n^2r_n + 12g_n^4 \cr\cr
& & + 52g_n^3h_n + 78g_n^2h_n^2 + 48g_nh_n^3 + 22h_n^4 + 6f_n^2r_ns_n + 14f_ng_n^2s_n + 28f_ng_nh_ns_n \cr\cr
& & + 14f_nh_n^2s_n + 120f_ng_nr_n^2 + 120f_nh_nr_n^2 + 88g_n^3r_n + 264g_n^2h_nr_n + 264g_nh_n^2r_n \cr\cr
& & + 88h_n^3r_n \ , 
\label{r3eq}
\eeqs
\beqs
s_{n+1} & = & 4f_ng_n^3 + 36f_ng_n^2r_n + 24f_ng_nh_nr_n + 12g_n^4 + 24g_n^3h_n + 12f_ng_n^2s_n + 24f_ng_nh_ns_n \cr\cr
& & + 12f_nh_n^2s_n + 144f_ng_nr_n^2 + 120f_nh_nr_n^2 + 144g_n^3r_n + 360g_n^2h_nr_n + 216g_nh_n^2r_n \cr\cr
& & + 144f_ng_nr_ns_n + 144f_nh_nr_ns_n + 56g_n^3s_n + 168g_n^2h_ns_n + 168g_nh_n^2s_n + 56h_n^3s_n \cr\cr
& & + 208f_nr_n^3 + 720g_n^2r_n^2 + 1440g_nh_nr_n^2 + 720h_n^2r_n^2 \ .
\label{s3eq}
\eeqs

\bibliographystyle{abbrvnat}

\end{document}